\documentclass[12pt]{article}
\usepackage[utf8]{inputenc}

\usepackage[top=0.85in,left=0.75in,right=0.75in, footskip=0.75in]{geometry}
\usepackage[utf8]{inputenc}
\usepackage{comment}
\usepackage{blindtext}
\usepackage{hyperref}
\usepackage{amsmath}
\usepackage{amssymb}
\usepackage{graphicx}
\usepackage{subfig}
\usepackage{amsfonts}
\usepackage{caption}
\usepackage{fancyhdr}
\usepackage{float}
\usepackage[none]{hyphenat}
\usepackage[greek,english]{babel}
\usepackage{afterpage}
\usepackage[utf8]{inputenc}
\usepackage[english]{babel}
\usepackage{color}
\usepackage{longtable}
\usepackage{lscape}
\usepackage{lipsum} % To generate text
\usepackage{array}
\usepackage{flafter}
\usepackage{color, colortbl}
\usepackage[pdftex,dvipsnames]{xcolor}  % Coloured text etc.
\usepackage{amsmath}
\usepackage[colorinlistoftodos,prependcaption]{todonotes}
\usepackage[numbers,sort&compress]{natbib}
\usepackage[pdftex,dvipsnames]{xcolor}  % Coloured text etc.
\usepackage{amsmath,bm}
\usepackage{xargs}                      % Use more than one optional parameter in a new commands
\usepackage[pdftex,dvipsnames]{xcolor} 
\usepackage{svg}
\usepackage{svg-extract}
\usepackage{booktabs,caption}

\title{Meta-analysis of diagnostic test accuracy with multiple disease stages: combining stage-specific and merged-stage data}
\author{Efthymia Derezea$^{1}$, Nicky J Welton$^{1}$, Gabriel Rogers$^{2}$, Hayley E Jones$^{1}$}
\date{February 2026}

\begin{document}

\maketitle
\begin{center}

	\begin{small}
		$1$. \textit{Population Health Sciences, Bristol Medical School, University of Bristol, UK}\\
		$2$. \textit{Manchester Centre for Health Economics, University of Manchester, UK}\\
	%	$2$.
	\end{small}	
	\end{center}
\begin{abstract}
For many conditions, it is of clinical importance to know not just the ability of a test to distinguish between those with and without the disease, but also the sensitivity to detect disease at different stages: in particular, the test's ability to detect disease at a stage most amenable to treatment. In a systematic review of test accuracy, pooled stage-specific estimates can be produced using subgroup analysis or meta-regression. However, this requires stage-specific data from each study, which is often not reported. Studies may, however, report test sensitivity for merged stage categories (e.g. stages I-II) or merged across all stages, together with information on the proportion of patients with disease at each stage. We demonstrate how such data can be modeled together with stage-specific data, to allow the inclusion of more studies in the meta-analysis. We consider both meta-analysis of tests with binary results, and meta-analysis of tests with continuous results, where the sensitivity to detect disease of each stage across the whole range of observed thresholds is estimated. The methods are demonstrated using a series of simulated datasets and applied to data from a systematic review of the accuracy of tests used to screen for hepatocellular carcinoma in people with liver cirrhosis. We show that incorporating studies with merged stage data can lead to more precise estimates and, in some cases, corrects biologically implausible results that can arise when the availability of stage-specific data is limited.  
\end{abstract}

\section{Introduction}\label{Sect:Intr}
Meta-analysis of diagnostic test accuracy combines all relevant evidence from multiple studies to summarize the accuracy of a diagnostic test, for a particular target condition \citep{Macaskill22,derezea2024technical}. Diagnostic test accuracy studies compare the test under evaluation to a reference standard that is usually assumed to be error-free \citep{trikalinos2012chapter}, with results reported in a $2\times 2$ contingency table showing the classification of patients into the diseased or disease-free category. Based on these results, we can calculate estimates of the accuracy of the index test, including the probability of correctly classifying a patient as diseased (sensitivity) or disease-free (specificity). Where tests produce a continuous, rather than binary, result, these $2\times 2$ tables may be reported based on dichotomizing results at different thresholds. In this situation, participants are considered to have tested positive if their test result is greater than a pre-specified threshold. Many studies report accuracy at multiple such thresholds. Binomial-bivariate \citep{reitsma2005bivariate,chu2006bivariate} or multinomial-multivariate meta-analysis models \citep{jones2019quantifying,steinhauser2016modelling} can be used to pool sensitivity and specificity across studies, for tests providing binary or continuous results respectively. \par 
Most meta-analysis methods focus on producing pooled accuracy estimates for a dichotomous health outcome (i.e. disease and disease-free). There are, however, many conditions --  including most types of cancer -- for which estimates of sensitivity for each of several stages of disease may be much more informative \citep{faria2018optimising,ng131}. Other examples include acute myocardial infarction \citep{haider2017high} and macular degeneration \citep{coleman2013age}.
 Estimates of the sensitivity of tests to detect disease at early stages, where it may be more amenable to treatment, may be particularly clinically relevant. Additionally, if evaluating tests for use in a screening context, a focus on overall sensitivity may be misleading, as the spectrum of disease in a symptomless screening population will often be shifted towards earlier stages. Consequently, `overall' sensitivity estimates, based on a wider spectrum of disease, may over-estimate sensitivity in this context. This `spectrum bias' is a common problem in both primary studies \citep{raffle2019screening} and systematic reviews of test accuracy \citep{lijmer1999empirical,ransohoff1978problems,lachs1992spectrum,whiting2004sources}, and may be mitigated by production of estimates of sensitivity stratified by disease stage \citep{lu2025examining}. \par  
We focus in this paper on the case where interest lies in the ability of a test to detect presence or absence of the condition (i.e. a binary classification), conditional on  the patient's true disease state and stage, rather than the test's ability to correctly classify patients into each disease stage. We revisit this distinction in the Discussion section.\par 
To produce meta-analytic estimates of sensitivity for each disease stage, ideally stage-specific estimates of sensitivity from each study would be used. In practice, however, availability of such data is often limited. While this data may be available from a -- potentially small -- subset of studies in a systematic review, other studies only provide what we will refer to as `merged stage' data, combining sensitivity across two or more stages. The most common form of merged stage data is overall sensitivity, i.e. sensitivity across all disease stages combined. Other studies may report merged sensitivity across a subset of stages, e.g. combined across stages I-II or across stages III-IV. Studies reporting merged stage sensitivity often also report the proportion of individuals at each stage of disease, as descriptive information. We will refer to these as `stage proportions' and to availability of this information alongside overall sensitivity as `overall' data. 
\par 

In this paper, we describe and demonstrate how overall and other merged stage data can be jointly modeled with stage-specific sensitivity data, and specificity, in a multivariate meta-analysis model, and explore whether incorporation of merged stage data increases precision of summary estimates. We  consider both meta-analysis of the accuracy of tests producing binary results, and  tests providing continuous results, where accuracy is reported across multiple thresholds in at least some of the studies. \par 
The remainder of this paper is structured as follows: Section \ref{Sect:Mot} presents a motivating example drawn from a systematic review of the accuracy of tests used to screen for hepatocellular carcinoma in people with liver cirrhosis. In Section \ref{Sect:Binary_model}, we formally define the problem for tests with binary results and describe the joint model for such tests. Section \ref{Sect:Cont_model} describes how the approach extends to accommodate tests with continuous results reported at various/multiple thresholds. In Sections \ref{sect:binary_applicA} and \ref{sect:cont_applic} the performance of the model for the binary and continuous results respectively is explored using artificial datasets and applied to data from the motivating example. Finally, Section \ref{Sect:Concl} concludes the paper and discusses potential directions for future work. 

\section{Motivating example: estimating sensitivity of tests for hepatocellular carcinoma by stage, in patients with cirrhosis}\label{Sect:Mot}
In a systematic review of the accuracy of tests in detecting hepatocellular carcinoma (HCC) in patients with liver cirrhosis \citep{rogers2025cirrhosis}, we aimed to produce meta-analyzed estimates of specificity and the sensitivity of each test to detect HCC at each stage. In addition to being of direct clinical relevance, stage-specific estimates of sensitivity are important parameters to inform subsequent health economic analyses. Since different studies reported stages using different staging systems, we first mapped these to approximate the Barcelona Clinic Liver Cancer (BCLC) staging system. Stage data were assigned to 3 BCLC groups: BCLC 0 (very early stage), BCLC A (early stage) and BCLC B/C/D (intermediate/advanced/terminal stage), which we will refer to throughout as `advanced' stage for brevity \citep{reig2022bclc}.  \par 

In principle, pooled estimates of sensitivity to detect each stage could be produced by performing meta-analysis of data relating to each stage separately, but very few studies reported this stage-specific data. For the five tests in the systematic review with stage-specific data available from at least one study, Table \ref{tab:hcc-data} summarizes the number of studies reporting data of each type. Of these five tests, data on four were reported as binary (HCC / no HCC), while one (AFP) is a continuous biomarker, for which accuracy data were reported at multiple thresholds. We see that, for each of the binary tests, only between 1 and 5 studies reported stage-specific sensitivity estimates. For one test, AFP-L3, the number of studies providing data for the `very early' and `early' stages - the most clinically important - was 2 and 1 respectively. We see, however, that there were 4 more studies reporting merged stage data for AFP-L3. 

For the continuous test, alpha-fetoprotein (AFP), a total of 110 studies were included in the systematic review,  reporting data on sensitivity and specificity across a wide range of diagnostic thresholds. This can be meta-analyzed to produce pooled estimates of sensitivity and specificity at all thresholds \citep{jones2019quantifying}. However, only 7 ($6\%$) and 6 ($7\%$) studies, respectively, reported data on the sensitivity of AFP to detect `very early' or `early' stage HCC. An additional 51 studies reported either `overall' (sensitivity across all disease stages and stage proportions) or other merged stage sensitivity data (sensitivity to detect `very early' or `early' stage HCC). This potentially contains useful information.\par

	\begin{table}[!h]
 		\centering
 		\caption{\label{tab:hcc-data}Summary of relevant data from HCC review. }
 		%\begin{centering}
 		%\centering
 		\resizebox{\textwidth}{!}{%
 			\begin{tabular}{lrrrrrr}
 				\multicolumn{2}{c}{} &
 				\multicolumn{5}{c}{Test*}\\ 
 				\textbf{}&\textbf{}&\textbf{US}&\textbf{MRI}	&\textbf{CT }&\textbf{AFP} &\textbf{AFP-L3 at 10\%} \\
 				\hline
                      & \textbf{Continuous test**} & no& no & no& yes& no \\
 				&\textbf{Stage-specific studies***}  & 50\% & 60$\%$&  71\%& 25\%& 67\%\\
                \hline
 				 Total studies / specificity data**** &\textbf{ No HCC (0)} & 16& 10 &7 &110 & 9\\
                 \hline
 				Studies reporting stage-specific data& \textbf{Very early (1)} &5 & 5 & 4& 7& 2\\
 				& \textbf{Early (2)} &4 &  3& 2& 6& 1\\
                & \textbf{Advanced (3)} &5 &  2& 1& 15& 5\\
                \hline
                Studies reporting & \textbf{Overall (Merged 1/2/3)} &4 &  2& 1& 30\textsuperscript{\textdagger}& 1\\
                & \textbf{Merged (1/2)} &3 &  2& 2& 21& 3
 				\\
 				\hline
 			\end{tabular}\hfill
 			%\end{centering}
            \vspace{0.5cm}
 		}
        
 {\raggedright \begin{small}\textit{* US=Ultrasound, MRI=Magnetic resonance imaging, CT=Computed tomography, AFP=Alpha-fetoprotein, AFP-L3=Lens culinaris agglutinin-reactive fraction of fetoprotein}\\
 \textit{** Accuracy data available at multiple thresholds per study}\\ \textit{*** Percentage of studies for each test that reported accuracy for at least one of the three cancer-stage stages.}\\ \textit{**** The number of ``no HCC" studies is equal to the total number of studies in the dataset for each test.}\\ \textsuperscript{\textdagger}\textit{This includes 19 studies reporting overall data as described in the main text, and 11 studies reporting a slightly different data structure - more details provided in Table \ref{tab:afp-data} and Supplementary Material.} \end{small}\par}
 	\end{table}

\begin{figure}[h]
    \centering
    \includesvg[width=300pt]{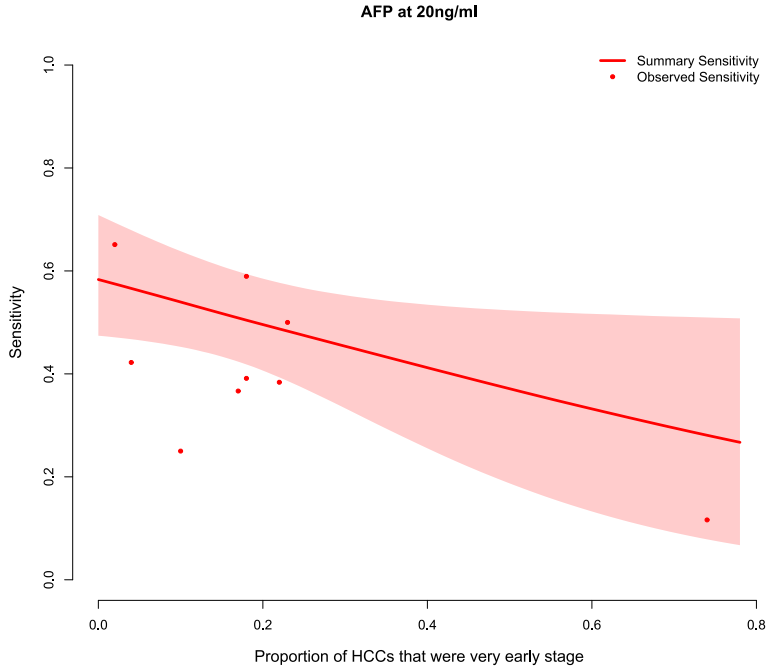}

    \caption{Results from a meta-regression of the sensitivity of AFP in detecting HCC, with the proportion of HCCs that were at a `very early' stage as a covariate, acting on the location parameter of the multiple thresholds model. The coefficient was estimated as -4.89 (95$\%$ CrI -9.94, -1.45). Data and results are shown  for the  threshold of 20ng/ml but the analysis is based on all available thresholds (a total of 31 data points). The shaded area represents the 95\% CrIs around estimated sensitivity.}
    \label{fig:metreg-afp}
\end{figure}

 One approach to investigating the relationship between sensitivity and HCC stage using the `overall' data would be to fit a meta-regression, regressing sensitivity on one or more of the stage proportions. For tests producing binary results, this can be done within the standard bivariate  \citep{reitsma2005bivariate,chu2006bivariate} or hierarchical summary receiver operating characteristic (HSROC) meta-analysis models \citep{rutter2001hierarchical}. Similarly, for continuous tests with accuracy reported at multiple thresholds in at least some studies, covariates can be incorporated into a multiple thresholds meta-analysis model  \citep{jones2019quantifying}. As an example, we fitted the multiple thresholds model of \citet{jones2019quantifying} to the 19 studies reporting `overall' data on the sensitivity of AFP, with the proportion of HCCs that were `very early' stage as a study-level covariate acting on the location parameter. This model uses data relating to - and produces pooled estimates of sensitivity and specificity for -  all thresholds, but for simplicity we show on Figure \ref{fig:metreg-afp} only the data and estimated sensitivity of AFP at the most commonly reported threshold, 20ng/ml. We see evidence of a relationship between sensitivity and the proportion of HCCs at a very early stage, with sensitivity reducing with higher proportions.  \par
%Note that these results are based on data from all available thresholds and not just the points shown in Figure \ref{fig:metreg-afp}. \par

This meta-regression suggests evidence of a relationship between HCC stage and AFP sensitivity, but is based only on relationships between - rather than within - studies, and does not directly provide estimates of the sensitivity conditional on HCC stage. Additionally, this analysis does not incorporate the clearly valuable information provided by the smaller number of studies reporting stage-specific sensitivity. In the following sections, we show how both sources of information can be modeled together, making the most of all available data to produce estimates of the sensitivity of each test to detect disease at each stage. 

\section{Model for binary test results}\label{Sect:Binary_model}
\subsection{Notation}\label{Sect:Binary_not}
Say that we wish to obtain estimates of the accuracy of a test for a target condition with $J+1$ disease states, indexed by $j=0$ (disease-free), $1,..., J$. For example, there are four states in total for our motivating dataset: `no HCC' ($j=0)$, and 3 stages of disease `Very early', `Early' and `Advanced' HCC $(j = 1,2,3)$.\par

Let $i = 1,...,I$ denote study number. We assume that all $I$ studies report data on stage $j = 0$, i.e. specificity. Ideally, each of the $I$ studies would report full stage-specific data as shown in Table \ref{tab:2jdata}, where $x_{ij}$ and  $x_{ij}^{-}$ represent the number of positive and negative test results, respectively, among individuals in each state $j$ in study $i$. The total number of participants in each state in each study is defined as $N_{ij}=x_{ij}+x_{ij}^{-}$. Note that $N_{ij}$ could be zero for some $j$, since not all studies will report data for all disease stages. \par

	\begin{table}[!h]
 		\centering
 		\caption{\label{tab:2jdata}Data format example for study $i$ reporting stage-specific accuracy data on $J+1$ disease states.}
 		%\begin{centering}
 		%\centering
 		%\resizebox{\textwidth}{!}{%
 			\begin{tabular}{lrrrrrr}
 				\multicolumn{2}{c}{} &
 				\multicolumn{5}{c}{True Disease State}\\  
 				 & &\textbf{Disease-free}&\textbf{Stage 1}	&\textbf{Stage 2}&\dots &\textbf{Stage $J$} \\
                & & {($j=0$)} & ($j = 1$) & ($j=2$) & \dots & ($j = J$)\\
                \hline
                    Test  & $+$ & $x_{i0}$&  $x_{i1}$& $x_{i2}$& \dots& $x_{iJ}$\\
 				Result  & $-$ &$x_{i0}^{-}$ &$x_{i1}^{-}$  &$x_{i2}^{-}$ & \dots& $x_{iJ}^{-}$
 				\\
                \hline
                Total  &  &$N_{i0}$ &$N_{i1}$  &$N_{i2}$ & \dots& $N_{iJ}$
 				
 			\end{tabular}\hfill
 			%\end{centering}
 		%}
 	\end{table}

In practice, however, only $I_{A}$ studies report stage-specific data. We assume that additional studies are available that report some collapsed version of the counts in Table \ref{tab:2jdata}. Taking our motivating dataset specifically as an example:
\begin{itemize}
    \item \textbf{`Overall' data}: An additional $I_{B}$ studies report $x_{i0}, x_{i0}^{-}, xall_i = \sum_{j=1}^{J}{ x_{ij}}$ and $xall_i^{-} = \sum_{j=1}^J{ x_{ij}^{-}}$. These studies also report the proportions of individuals within each of the $J$ disease stages, defined as $p_{ij}=\frac{N_{ij}}{\sum_{j=1}^{J}N_{ij}}$.
    \item \textbf{Other merged data}: A further $I_{C}$ studies report $x_{i0}$ and $x_{i0}^{-}$ and other merged sensitivity data. In our motivating example, these studies reported sensitivity combined over `very early' and `early' stages, and also sensitivity for `advanced' stage, i.e.  $xm_i=x_{i1}+x_{i2}$, $xm_i^{-}=x_{i1}^{-}+x_{i2}^{-}$, $x_{i3}$ and $x_{i3}^{-}$. Note that, for this particular example,  the proportion of individuals across the merged disease stages that belonged to the first disease stage, $q_i=\frac{p_{i1}}{p_{i1}+p_{i2}}$, was unknown, although this may also be reported in other examples.
\end{itemize}
Note that in the case of studies reporting specificity and overall sensitivity without stage proportions, only the disease-free data are included in the analyses presented in this paper. We denote the number of these specificity-only studies as $I_{D}$. The total number of studies is then $I = I_{A} + I_{B} + I_{C} + I_{D}$. 

\subsection{Model for complete stage-specific data }\label{sect:jvar}
If all studies reported the ideal, stage-specific, data, or if we chose to model only the data from studies that reported this, we could produce summary estimates of the probability of a positive test result for individuals in each state using subgroup analysis or meta-regression \citep{Macaskill22}. Where the index test is binary, a natural approach might be to extend the bivariate random effects meta-analysis for sensitivity and specificity \citep{reitsma2005bivariate,chu2006bivariate} to include a covariate for disease stage, acting on sensitivity. However, this would involve an implicit assumption of perfect correlation between studies in the sensitivity to detect disease of each stage, and constant between-study heterogeneity.  \par

 To allow for a more general covariance structure, a natural extension of the bivariate model is a $(J+1)$-variate model, where the number of people in each disease state testing positive follows a conditionally independent Binomial distribution,
\begin{equation}\label{eq:bin_jvar}
    x_{ij} | S_{ij} \sim \mathrm{Binomial}(S_{ij},N_{ij}),\quad j=0,\dots,J
\end{equation}
Here, $S_{i0}$ is the false positive fraction (FPF=1-Specificity) in study $i$, and $S_{i1},\dots,S_{iJ}$ denote the corresponding sensitivities of the test conditional on each disease stage. For example, for the HCC data set $S_{i1} = \Pr(\text{test positive} \mid \text{very early stage HCC})$.\par
As in the bivariate model, we assume that, for some monotonic transformation $g:(0,1)\rightarrow \mathbb{R}$, the $g-$transformed probabilities follow a multivariate normal distribution of the form:
\begin{equation}\label{eq:res_jvar}
\begin{bmatrix}
    g(S_{i0})    \\
    \vdots \\
    g(S_{iJ})    \\
    
    \end{bmatrix}
 \sim N\left(
\begin{bmatrix}
    m_{0}  \\
    \vdots \\
    m_{J}
    \end{bmatrix}
,\mathbf{\Sigma} \right)
\end{equation}
where with $m_j$ we denote the mean of the g-transformed probability of testing positive among those in the $j$th state. This is a simple extension of the standard bivariate model \cite{reitsma2005bivariate,chu2006bivariate}, moving from two disease states (diseased and disease-free) to $J+1$ states. The $(J+1)\times (J+1)$ variance-covariance matrix, $\mathbf{\Sigma}$, allows for correlations across studies in these probabilities, which can be anticipated to arise due to study-level factors. For example, in studies with a relatively high sensitivity for the very early stage, we might expect to observe higher sensitivities for the other disease stages too, due to other factors such as how the test was conducted. Observations $x_{ij}$ within each study are independent, conditional on $S_{ij}$, since they are based on test results in different groups of individuals. \par 

Throughout this paper, we adopt the logit function for the g-transform. We obtain pooled estimates of sensitivities and specificity by calculating the inverse logit of the parameters $m_j$ as shown below:
$$\operatorname{\textrm{Summary Specificity}}=1-\frac{\exp(m_{0})}{1+\exp(m_0)}$$
$$\operatorname{\textrm{Summary S}\mathrm{ensitivity_j}}=\frac{\exp(m_{j})}{1+\exp(m_j)},\quad j=1,\dots,J$$

\subsection{Joint model for stage-specific and merged stage data}\label{sect:binary_full}
When stage-specific data are available from some studies but only merged stage data from others, we can jointly model data from all studies. The model for the stage-specific data from studies $i = 1,..., I_A$ is characterized as described above (Section \ref{sect:jvar}). Note that for the disease-free state the likelihood can informed by all $I$ studies and not just the $I_{A}$ that report stage-specific data.\par 

Simultaneously, the overall count, $xall_i,\quad i=I_A +1,\dots,I_A +I_{B}$, from each study reporting stage proportions ($j=1,\dots,J$), has the following Binomial likelihood,
$$xall_{i}|S_{i1},\dots,S_{iJ}\sim \mathrm{Binomial}\left(\sum_{j=1}^{J}p_{ij}S_{ij},\sum_{j=1}^{J}N_{ij}\right)$$ 
where the overall sensitivity is a weighted average of the sensitivities across all $J$ stages, with weights $p_{ij}$. \par

Similarly, the counts $xm_i, \quad i=I_A +I_B +1,\dots,I_A+I_B+I_C$ of the merged stages 1 and 2 are modeled as,
$$xm_i|S_{i1},S_{i2} \sim \mathrm{Binomial}\left(q_{i}S_{i1}+(1-q_i)S_{i2},N_{i1}+N_{i2}\right)$$ 
with the sensitivity of the merged stage being the weighted average of the sensitivities across stages 1 and 2. Note that in this case, $N_{i1}$ and $N_{i2}$ (and hence $q_i$) are not known, and only their summation is available. \par

The between-studies model is the same as described in Section \ref{sect:jvar}, but is now jointly estimated from all four types of studies.

\section{Model for continuous test results}\label{Sect:Cont_model}
\subsection{Notation}
We now demonstrate how we can obtain accuracy estimates for each disease state across the whole range of observed thresholds, by extending the joint modeling approach described in Section \ref{Sect:Binary_model} to the multiple thresholds model by \citet{jones2019quantifying}.\par 

We extend the notation introduced in Section \ref{Sect:Binary_not} by including a further index $t=1,\dots,T_{ij}$, indicating the specific threshold $C_{ijt}$ that each reported data point corresponds to, where $T_{ij}$ is the number of thresholds that study $i$ reports for disease state $j$. Table \ref{tab:mult_jdata} shows the ideal data each study would provide. Here, $x_{ijt}$ and $x_{ijt}^{-}$ represent the number of participants with a test result larger and smaller than $C_{ijt}$ respectively. As previously, $N_{ij}$ denotes the total number of participants in each study and state. Again, $N_{ij}$ can be zero if a study does not report on all disease states. Assuming that higher values of the continuous test result in a higher probability of disease then by definition, within each study and disease state, as the threshold $C_{ijt}$  increases, $x_{ijt}$ must decrease or remain the same. 

	\begin{table}[!h]
 		\centering
 		\caption{\label{tab:mult_jdata}Data format example for study $i$ reporting accuracy on $J+1$ disease states at multiple thresholds.}
 		%\begin{centering}
 		%\centering
 		%\resizebox{\textwidth}{!}{%
 			\begin{tabular}{lr|rrr|rrr|r|rrr}
 				\multicolumn{2}{c}{} &
 				\multicolumn{10}{c}{True Disease State}\\ 
                     \multicolumn{2}{c}{} &\multicolumn{3}{c}{\textbf{Disease-free}}&\multicolumn{3}{c}{\textbf{Stage 1}}&\multicolumn{1}{c}{...}&\multicolumn{3}{c}{\textbf{Stage J}}\\
                     \multicolumn{2}{c}{} &\multicolumn{3}{c}{\textbf{($j=0$)}}&\multicolumn{3}{c}{\textbf{($j=1$)}}&\multicolumn{1}{c}{...}&\multicolumn{3}{c}{\textbf{($j=J$)}}\\
                     \hline
 				 \multicolumn{2}{c}{\textbf{Threshold}} &\multicolumn{1}{c}{$C_{i01}$}&\multicolumn{1}{c}{...}&\multicolumn{1}{c}{$C_{i0T_{i0}}$}&\multicolumn{1}{c}{$C_{i11}$}&\multicolumn{1}{c}{...}&\multicolumn{1}{c}{$C_{i1T_{i1}}$}&\multicolumn{1}{c}{...}&\multicolumn{1}{c}{$C_{iJ1}$}&\multicolumn{1}{c}{...}&\multicolumn{1}{c}{$C_{iJT_{iJ}}$} \\
 				\hline
                    Test  & $+$ & $x_{i01}$&\dots  &$x_{i0T_{i0}}$ &$x_{i11}$ &\dots &$x_{i1T_{i1}}$ &\dots &$x_{iJ1}$ &\dots &$x_{iJT_{iJ}}$ \\
 				Result  & $-$ &$x_{i01}^{-}$&\dots  &$x_{i0T_{i0}}^{-}$ &$x_{i11}^{-}$ &\dots &$x_{i1T_{i1}}^{-}$ &\dots &$x_{iJ1}^{-}$ &\dots &$x_{iJT_{iJ}}^{-}$
 				\\
                \hline
                Total  &  & &$N_{i0}$  & & &$N_{i1}$ & &\dots & &$N_{iJ}$ &
 				
 			\end{tabular}\hfill
 			%\end{centering}
 		%}
 	\end{table}

In a similar manner, we extend the notation for the studies that report sensitivity for merged disease states. For an `overall' study $i$ with stage proportions, we denote by $xall_{it}$ and $xall_{it}^{-}$, $t=1,\dots,Tall_i$, the number of participants across all stages of disease (i.e., $\sum_{j=1}^{J}x_{ijt}$) with a test result larger and smaller than the $t$th threshold, respectively. $Tall_i$ denotes the number of thresholds reported at in study $i$, and the total number of diseased is defined as $Nall_{i}= \sum_{j=1}^{J}  N_{ij} $. \par

For the AFP test in our motivating example, there were two other types of merged data available. Details are provided in the Supplementary Material.

\subsection{Model for complete stage-specific data}\label{sect:cont_sub}
As with the binary test result model, we consider first the situation where either all studies report the ideal stage-specific data, or we choose to model only this type of data. As described by \citet{jones2019quantifying}, counts $x_{ijt}$ follow a series of conditional Binomial distributions: 
$$x_{ij1}|S_{ij1} \sim \mathrm{Binomial}(S_{ij1},N_{ij})$$
\begin{equation}\label{eq:cond_binom}
    x_{ijt}|x_{ij,t-1},S_{ijt}, S_{ij,t-1}\sim \mathrm{Binomial}\left(\frac{S_{ijt}}{S_{ij,t-1}},x_{ij,t-1}\right),\quad t=2,\dots,T_{ij}
\end{equation}

where $S_{ijt}$ is the probability of a positive test result in the $i$th study, $j$th disease state for the $t$th threshold. Under the assumption that a monotonic function $f(\cdot)$ transforms the distribution of test results in each disease state to a Logistic, with location and scale parameter $\mu_{ij}$ and  $\sigma_{ij}$ respectively, it follows that
$$\mathrm{logit}(S_{ijt})=\frac{\mu_{ij}-f(C_{ijt})}{\sigma_{ij}}$$
Note that $f(\cdot)$ can be any monotonically decreasing function, for example, any transformation from the Box-Cox family:
$$f(C_{ijt})= \begin{cases} 
      \frac{C_{ijt}^{\lambda}-1}{\lambda} & \lambda \neq 0 \\
      \log(C_{ijt}) & \lambda = 0 
   \end{cases}
$$
where $\lambda$ can be either pre-defined (e.g. set to zero) or estimated from the data as an additional model parameter. \par 
For the standard two disease-state case, \citet{jones2019quantifying} described a quadrivariate distribution for the location and (log) scale parameters, although also discussed simplifications of this dependence structure. A natural extension of this between-studies model to $J+1$ disease states is a $2(J+1)$-variate normal distribution,

\begin{equation}\label{eq:re_2J}
    \begin{bmatrix}
    \mu_{i0}    \\
    \vdots\\
    \mu_{iJ}\\
    log(\sigma_{i0})    \\
    \vdots\\
    log(\sigma_{iJ})
    \end{bmatrix}
 \sim N\left(
\begin{bmatrix}
    m_1  \\
    \vdots\\
    m_{2(J+1)}
    \end{bmatrix}
,\mathbf{\Sigma_t} \right)
\end{equation}
where $\mathbf{\Sigma_t}$ is an unrestricted $2(J+1)\times 2(J+1)$ variance-covariance matrix and $m_1,\dots,m_{2(J+1)}$ are the random-effect means of the location and log-transformed scale parameters. The summary specificity and sensitivity per stage estimates, for any threshold $C$ within the observed threshold range, are given by: 
$$\operatorname{\textrm{Summary Specificity}}(C)=1-\mathrm{logit}^{-1}\left(\frac{m_{1}-f(C)}{e^{m_{J+2}}}\right)$$
$$\operatorname{\textrm{Summary S}\mathrm{ensitivity_{j}}}(C)=\mathrm{logit}^{-1}\left(\frac{m_{j+1}-f(C)}{e^{m_{j+J+2}}}\right),\quad j=1,\dots,J$$

\subsection{Joint model for stage-specific and merged stage data}\label{sect:overall_cont}
To model jointly both stage-specific and merged stage data, the model for the stage-specific data from studies $i = 1,...I_A$ is as specified above. Note that, similarly to the binary test case, the part of the likelihood corresponding to the disease-free state can be informed by all available studies and not just the $I_A$ ones. \par

The `overall' counts $xall_{it}$, $i=I_A +1,\dots,I_A+I_B$, follow a series of conditional Binomial distributions,
$$xall_{i1}|S_{i11},\dots,S_{iJ1}\sim Bin\left(\sum_{j=1}^{J}p_{ij}S_{ij1},Nall_{i}\right)$$
$$xall_{it}|xall_{i,t-1},S_{i1t},\dots,S_{iJt},S_{i1,t-1},\dots,S_{iJ,t-1}\sim Bin\left(\frac{\sum_{j=1}^{J}p_{ij}S_{ijt}}{\sum_{j=1}^{J}p_{ij}S_{ij,t-1}},xall_{i,t-1}\right),\quad t=2,\dots,Tall_{i}$$
The between-studies model is given by \eqref{eq:re_2J}, shared with the model for stage-specific data, such that parameters $m_1,\dots,m_{2(J+1)}$, $\mathbf{\Sigma_t}$, and $\lambda$ are jointly informed by both sources of data. \par

Details of how we incorporated studies reporting the two other types of merged data in our motivating example into this joint model are available in the Supplementary Material.

\section{Application of the binary test result model }\label{sect:binary_applicA}
\subsection{Analysis of artificial data}\label{sect:binary_sims}

To explore whether there is potential for the `overall' type of data to improve inference when limited stage-specific data are available, we simulated artificial data from 20 studies, under the following assumptions. First we assume the data represent 4 different disease states, i.e. $J=3$ plus the disease-free state, with the true average values across studies of the probabilities of a positive test result being 0.05, 0.70, 0.80, and 0.95 respectively, i.e., specificity of 95\% and sensitivities of 70\%, 80\% and 95\% for each of the 3 disease stages. We assumed a positive correlation between each of the $J+1$ random effects of magnitude 0.3, and variances of 0.05, 0.2, 0.3 and 0.6, representing a realistic degree of heterogeneity across studies. \par

We consider the situation in which only 4 of the 20 studies contribute stage-specific data, while the remaining 16 studies provide `overall' data. We simulated the data at the stage level for all 20 studies, prior to aggregation in 16 studies to replicate this hypothetical scenario. We generated data corresponding to three scenarios with different levels of `noise', defined by the magnitude of $var_{ij}$, and how this noise varied across these two types of study, where $var_{ij}=N_{ij}S_{ij}(1-S_{ij})$, is the variance of the binomial data. We set $var_{ij}$ and solved with respect to sample size, $N_{ij}$. For scenario (a) we set $var_{ij} = 0.5$ for both types of data. Scenario (b) assumes a smaller level of noise for the stage-specific data (with variances ranging from 0.3 to 0.5) compared with the `overall' data (with variances between 0.9 and 1.2). Scenario (c) assumes the opposite, with smaller variances for the `overall' data than for the stage-specific data. Finally, we simulated counts $x_{ij}$ from the model defined in Section \ref{sect:jvar}. The code for generating the artificial data described and reproducing the following analyses is available on \href{https://github.com/FeniaDerezea/MDS_MA_DTA}{Git-Hub}. \par 

 To replicate the scenario in which stage-specific data are only available for 4 studies, we aggregated the counts across the three disease stages (excluding the `disease-free' state) for the 16 `overall' studies. We then modeled three different versions of the simulated dataset: (1) the ideal full stage-specific  data from all 20 studies, using the quadrivariate model introduced in Section \ref{sect:jvar}; (2) only the hypothetically observed stage-specific data from the four studies, again using the quadrivariate model; and (3) the full observed hypothetical dataset comprising of both stage-specific and overall data, using the joint model described in Section \ref{sect:binary_full}. We compared summary estimates of sensitivity and specificity across the three analyses, treating the results from version (1) as the `gold standard'. \par 

Throughout, we operate within a Bayesian framework,  using JAGS \citep{plummer2004jags} as an MCMC sampler called through R using package R2jags \citep{su2015package}. To complete the model, we assign priors $m_j\sim \mathrm{Logistic(0,1)}$ and $ \mathbf{\Sigma} \sim \mathrm{Inv.Wishart}(R,4)$, where $R$ is a $4\times4$ identity matrix. These priors for $m_j$ are equivalent to assigning vague $\mathrm{Uniform}(0,1)$ priors on the probability scale \citep{derezea2024technical}.

	\begin{table}[!h]
 		\centering
 		\caption{\label{tab:binary_sim}Summary estimates (with 95$\%$ CrIs) of the probability of a positive test conditional on disease state, from meta-analysis of artificial data. }
 		%\begin{centering}
 		%\centering
 		\resizebox{\textwidth}{!}{%
 			\begin{tabular}{lr|rrrr}
 				\multicolumn{2}{c}{} &
 				\multicolumn{4}{c}{State}\\  
 				\textbf{ }&\textbf{ }&\textbf{Disease-free}&\textbf{Stage 1}&\textbf{Stage 2}&\textbf{Stage 3}\\
 				\hline
                    \textbf{} & Truth &0.05  & 0.70 & 0.80  & 0.95  \\
                    \hline
                    \textbf{} &\textbf{(1) Ideal} & 0.05 (0.02,0.09)&0.58 (0.38,0.75)&0.84 (0.71,0.93)&0.95 (0.91,0.98) \\
 				Scenario (a)& \textbf{(2) Observed stage-specific}&0.05 (0.02,0.09)&0.54 (0.18,0.86)&0.79 (0.36,0.96)&0.94 (0.71,0.99)  \\
                \textbf{} & \textbf{(3) Observed joint}& 0.05 (0.02,0.09)&0.53 (0.24,0.82)&0.83 (0.57,0.96)& 0.97 (0.91,0.99)\\
                \hline 
 				\textbf{}& \textbf{(1) Ideal} &0.04 (0.02,0.07)&0.63 (0.47,0.76)&0.84 (0.73,0.92)&0.96 (0.92,0.98)  \\
                Scenario (b)&\textbf{(2) Observed stage-specific} &0.04 (0.02,0.07)&0.66 (0.23,0.93)& 0.78 (0.35,0.96)&0.95 (0.72,0.99)\\
 				\textbf{}&\textbf{(3) Observed joint} &0.04 (0.02,0.07)&0.62 (0.31,0.89)&0.80 (0.53,0.95)&0.98 (0.93,1.00)  \\
                \hline 
                \textbf{}& \textbf{(1) Ideal}&0.05 (0.02,0.09)&0.57 (0.33,0.76)&0.81 (0.65,0.92)&0.95 (0.90,0.98)\\
                Scenario (c)& \textbf{(2) Observed stage-specific}& 0.05 (0.02,0.09)&0.50 (0.16,0.81)& 0.74 (0.33,0.93)&0.94 (0.73,0.99)\\
                 & \textbf{(3) Observed joint} &0.05 (0.02,0.10)& 0.55 (0.25,0.81)& 0.79 (0.53,0.93)&0.96 (0.90,0.99)
 				\\
 			\hline
 			\end{tabular}\hfill
 			%\end{centering}
            \vspace{0.5cm}
 		}
        {\raggedright \begin{small}\textit{* Scenario (a)= Same level of `noise' for stage-specific and overall data, Scenario (b)= Smaller level of `noise' for stage-specific data compared to overall, Scenario (c)= Smaller level of `noise' for overall data compared to stage-specific.}\\ \textit{**Rows `(1) Ideal' show results from fitting the quadrivariate model to stage-specific data from all studies (some of which is not available in practice). Rows `(2) Observed stage-specific' show results from the quadrivariate model applied to only the observed stage-specific data. Rows `(3) Observed Joint' show results from the joint model applied to the observed stage-specific and `overall' studies. }\end{small}\par}
 	\end{table}
   
Table \ref{tab:binary_sim} shows the summary estimates of the probability of a positive test conditional on each disease state, with 95$\%$ CrIs, obtained for the nine combinations of different scenarios and versions of the dataset used. The contributions to the disease-free state are the same across all analyses, for each scenario, and so we would not expect any precision gains for this state. We see, however, that by incorporating in the synthesis the information from the `overall' data, we obtain more precise estimates for the diseases stages, in all cases - even for scenario (b), where the additional overall data are more noisy compared with the observed stage-specific data. The increase in precision is greatest in scenario (c), where the `overall' studies provide the most precise data. For this particular example, the increase in precision is more obvious across all three scenarios for disease stages 2 and 3. Some of the improvements in precision are potentially meaningful for inference - in particular, we note that the 95\% CrIs around sensitivity to detect stages 1 and 3 of the disease no longer overlap after incorporating the `overall' data.

\subsection{Analysis of binary test data from the motivating example}\label{sect:binary_applic}
We now apply the joint model defined in Section \ref{sect:binary_full} to the four binary tests from the motivating example, and compare with results from meta-analyzing only the available stage-specific data using the quadrivariate model described in Section \ref{sect:jvar}. \par

 As described in Section \ref{Sect:Mot}, the motivating example data set contained both `overall' and other merged-stage studies, and the proportion of individuals from the merged(1/2) group that belong to the `very early' stage, $q_i$ is not known. To complete the model in this case we treated the $q_i$'s as nuisance parameters to be estimated from the data, with prior distribution $q_i \sim \mathrm{Uniform}(0,1)$.\par 

\begin{figure}[h]
    \centering
    \includesvg[width=500pt]{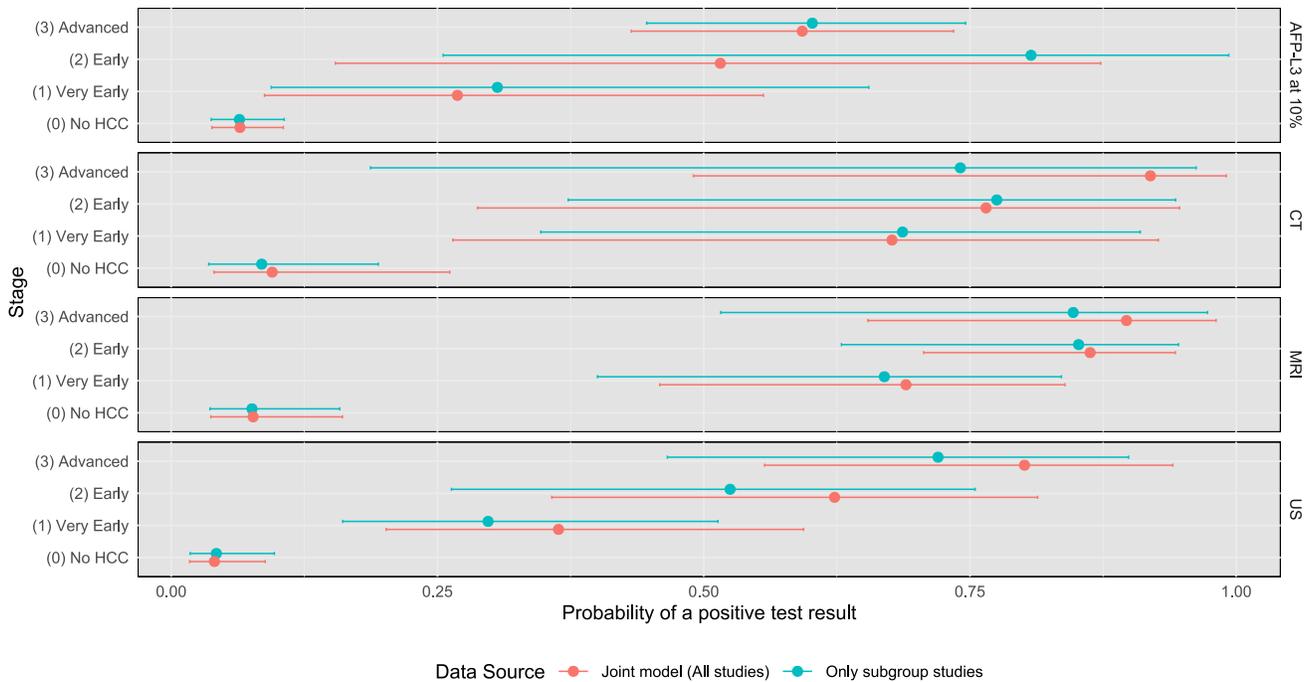}

    \caption{Accuracy of tests to detect HCC in people with cirrhosis. The circles represent summary estimates of the probability of a positive test result and the bars around them are 95\% CrIs. (US=Ultrasound, MRI=Magnetic resonance imaging, CT=Computed tomography, AFP=Alpha-fetoprotein, AFP-L3=Lens culinaris agglutinin-reactive fraction of fetoprotein) }
    \label{fig:binary-apll}
\end{figure}

The summary estimates with 95\% CrIs from these analyses are shown in Figure \ref{fig:binary-apll}. We notice that when only the studies reporting stage-specific data are used, the point estimates of sensitivity for the tests MRI, CT and AFP-L3 for the `early' stage appear larger than those for the `advanced' stage (although with considerable overlap between CrIs). This result seems implausible and is almost certainly caused by sampling error inherent in the very small amount of available stage-specific data. After incorporating the additional information from the `overall' studies into the analysis, this anomalous result is no longer observed. All point estimates from the joint model are consistent with sensitivity increasing as cancer stage advances, as expected, although 95\% CrIs still overlap. \par 

As expected, the FPF estimates (results for the disease-free state) are almost identical between the two models (with the exception of CT where we notice a small discrepancy) since all studies reported data for the disease-free state, such that there is no additional information in the joint model. Finally, we notice a small increase in precision in most cases when the full data is used, although - because of the very small amount of available data in total (see Table \ref{tab:hcc-data} for more details) - these differences are not striking in this case. An exception to this is the sensitivity of CT to detect advanced HCC: here, the 95\% CrI from the joint model is much narrower than that from the stage-specific data alone, although still wide.  \par

\section{Application of the continuous test results model}\label{sect:cont_applic}
\subsection{Analysis of artificial data}\label{sect:cont_sims}
We simulated two artificial data examples to explore potential gains in estimation when all available data is used, in the situation where the accuracy of a continuous test is reported at multiple thresholds per study (data available on \href{https://github.com/FeniaDerezea/MDS_MA_DTA}{Git-Hub}). In both cases, we imagine a continuous test producing results in the range $[5,150]$. We generated data corresponding to 30 studies and 4 disease states. We assume that each study reported between 1 and 10 thresholds. The number of thresholds $T_{ij}$ reported in each study was sampled from a $Uniform[1,10]$ distribution, and the numerical value of the $t^{th}$ threshold reported on in study $i$, $C_{it}$ was sampled from a $Uniform[5, 150]$ distribution.  We set $m_1,\dots,m_{2(J+1)}$, defined in equation \eqref{eq:re_2J}, as (0.81, 1.62, 2.56, 5.20, 0.19, 0.25, 0.39, 0.43). Table \ref{tab:cont_sim} shows the implied (summary) probabilities of testing positive for individuals in each disease state, for two example thresholds. \par

We assigned an $8\times8$ variance-covariance matrix for the random effects distribution, with the variances on the diagonal being (0.1, 0.2, 0.3, 0.4, 0.01, 0.05, 0.08, 0.1), reflecting reasonable heterogeneity across studies. We assumed equal and relatively strong correlations of 0.6 between the location parameters $\mu_{ij}$ for the four states. The correlations between the log-scale parameters and those between location and log-scale of the four states were assumed to be weaker, and given the values of 0.1 and 0.15 respectively. Sample sizes $N_{ij}$ were selected randomly from the range of $[200,700]$ for the disease-free group and range $[5,50]$ for the disease stages. Finally, for simplicity, a log function was adopted for the test result transformation $f$. The data were then simulated using equations \eqref{eq:cond_binom} and \eqref{eq:re_2J} for all studies.\par 

Similarly to Section \ref{sect:binary_sims}, we recreated a scenario in which the simulated stage-specific data are observed for only five of the 30 studies, with only `overall' data for the remaining 25 studies. To create this overall data, we aggregated the counts across the three disease stages for 25 of the studies at each threshold, i.e., $xall_{it}=\sum_{j=1}^{J}x_{ijt}, \quad i=6,\dots,30$. 
\par

For the first example, we assume that multiple threshold data are available for only the disease-free and Stage 3 states (Scenario I). For the second example, we consider the case where stage-specific multiple threshold data is available for all four disease states (Scenario II). \par

 As in our simulation for binary tests, we analyze three different versions of the data: (1) the ideal data (stage-specific data from all 30 studies), to which we fit the model defined in Section \ref{sect:cont_sub}; (2) the `observed' stage-specific data (from only 5 studies), to which we fit the \citet{jones2019quantifying} model independently for each state; and (3) the `observed' stage-specific (5 studies) and `overall' data (25 studies) modeled jointly as proposed in Section \ref{sect:overall_cont}. We compare the summary accuracy estimates obtained according to the three different versions across all thresholds. \par 
 We complete the models by assigning prior distributions
$m_j \sim N(0,100)$ for $\quad j=1,\dots,8$ and $\mathbf{\Sigma_t} \sim \mathrm{Inv.Wishart}(R,8)$. 
Note that, if the more flexible version of the model with unknown Box-Cox transformation parameter, lambda, is used, then a prior for that parameter is also needed, e.g. $\lambda \sim \mathrm{Uniform}(-3,3)$. For this example we assume that $\lambda$ is known and equal to 0.\par
\begin{figure}[h]
    \centering
    \includesvg[width=500pt]{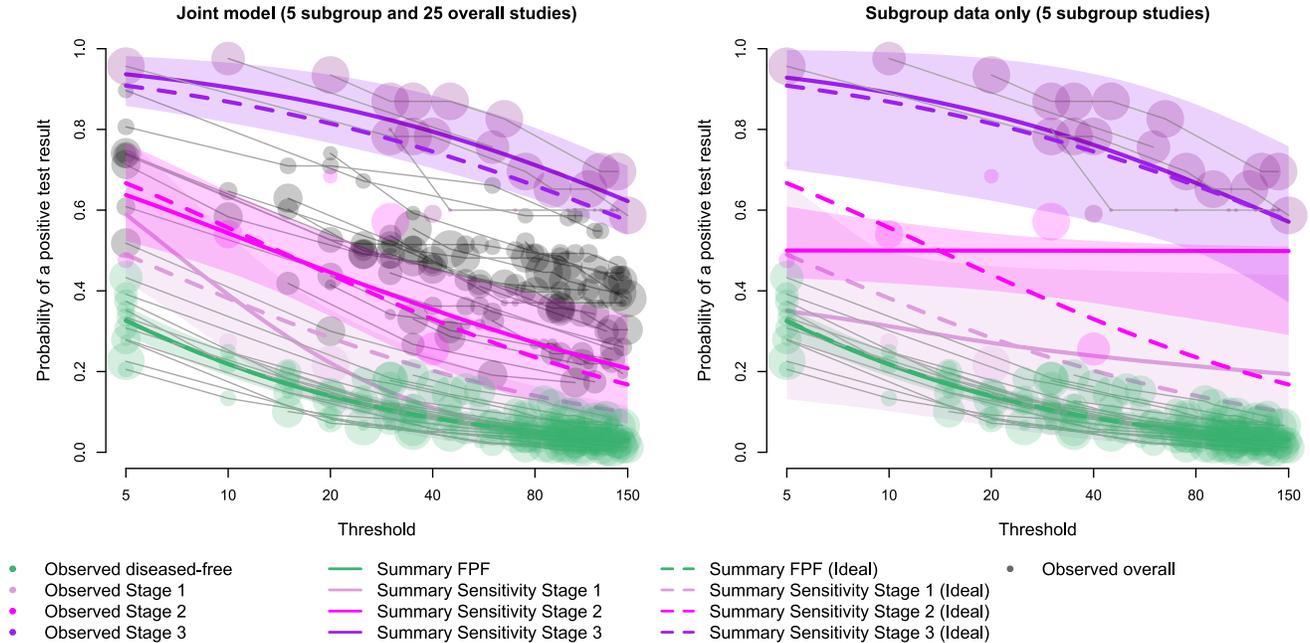}

    \caption{Results from models fitted to artificial data, Scenario I. Solid lines show results from fitting models to incomplete data. Dashed lines (identical across the two plots) show results from fitting models to the ideal full data, if it were available, and are shown for comparison. Shaded areas represent 95\% CrIs around the point estimates. The circles representing the observed points vary in diameter according to the sample size. Points connected by a gray line represent data belonging to the same study. FPF=False positive fraction.}

    \label{fig:sims-cont-s1}
\end{figure}

 
For Scenario I, the generated data are shown in Figure \ref{fig:sims-cont-s1}. Results from the joint model using both types of data are shown in the left panel and results using only the `observed' stage-specific data in the right panel of Figure \ref{fig:sims-cont-s1}. Unsurprisingly, when only the stage-specific data is used, the model struggles to estimate the slopes corresponding to the disease stage 1 and stage 2 states (which did not have multiple threshold data). This is especially striking for disease stage 2, where results suggest no changes in sensitivity as threshold increases, which seems unlikely. In contrast, when the joint model is used, making use of all available information, we are able to estimate a slope and obtain more intuitive results for all disease states. We see that the summary results for disease stages 1 and 2 are much closer to the results based on the ideal (largely unobserved) data when the additional information from the overall studies is utilized. Additionally, we observe a gain in precision of summary results across all disease states. \par

 Since this is not a full simulation study and we only generated a single data set, with sampling error, we would not expect to fully recover the `truth'. However, we might consider analysis of the ideal data set as the `gold standard' analysis for this data set. These are the results we would expect to obtain for this particular dataset had we observed all of the stage-specific data. Table \ref{tab:cont_sim} provides a comparison of results from modeling the ideal data with results from modeling only the stage-specific or the stage-specific plus overall data, for two arbitrarily chosen thresholds. As expected, the results for the disease-free group are identical in all cases, since exactly the same information was used across all three models. For the three disease stages, we notice that the point estimates of sensitivity for both thresholds obtained using the joint model, are in almost all cases closer to the ones based on the ideal/unobserved data compared with those obtained using only the observed stage-specific data. In all cases, the corresponding 95$\%$ CrIs from the joint model are in closer agreement with those using the ideal dataset and narrower than those based on the limited `observed' stage-specific data.

	\begin{table}[!h]
 		\centering
 		\caption{\label{tab:cont_sim}Summary estimates of the probability of a positive test with 95$\%$ CrIs for two thresholds}
 		%\begin{centering}
 		%\centering
 		\resizebox{\textwidth}{!}{%
 			\begin{tabular}{lrr|rrr|rrr}
 				\multicolumn{3}{c}{} &
 				\multicolumn{3}{c}{Scenario I}&\multicolumn{3}{c}{Scenario II }\\  
 				\textbf{Threshold}&\textbf{State}& \textbf{Truth}&\textbf{Observed stage-specific}&\textbf{Joint}&\textbf{Ideal stage-specific}	&\textbf{Observed stage-specific}&\textbf{Joint} &\textbf{Ideal stage-specific} \\
 				\hline
                     &\textbf{Disease-free}&0.23 &  0.22 (0.19,0.25) &  0.22 (0.19,0.25)&  0.22 (0.19,0.25) &  0.22 (0.19,0.25) & 0.22 (0.19,0.25) &  0.22 (0.19,0.25)\\
 				10&\textbf{Disease stage 1}& 0.37&0.31 (0.10,0.50)& 0.41 (0.28,0.54)& 0.38 (0.33,0.44)&  0.31 (0.12,0.50) & 0.38 (0.27,0.50)& 0.38 (0.32,0.44)\\
 				&\textbf{Disease stage 2} &0.54 & 0.50 (0.42,0.57)& 0.52 (0.43,0.62)& 0.56 (0.49,0.62)&  0.56 (0.38,0.79)&  0.54 (0.44,0.65)& 0.55 (0.49,0.62)\\
 				&\textbf{Disease stage 3}& 0.87& 0.89 (0.67,0.99)& 0.90 (0.81,0.96)& 0.87 (0.82,0.91)& 0.89 (0.67,0.99)& 0.90 (0.81,0.96)&  0.87 (0.82,0.91)\\
                \hline
                 &\textbf{Disease-free} & 0.04&  0.04 (0.03,0.06) & 0.04 (0.03,0.06)&  0.04 (0.03,0.06) & 0.04 (0.03,0.06) & 0.04 (0.03,0.06) &  0.04 (0.03,0.06)\\
 				100&\textbf{Disease stage 1}&0.09 & 0.21 (0.01,0.44)& 0.04 (0.01,0.14)& 0.12 (0.09,0.17)&  0.10 (0.01,0.27) & 0.08 (0.02,0.18)& 0.12 (0.09,0.17)\\
 				&\textbf{Disease stage 2} &0.20 & 0.50 (0.32,0.51)&  0.25 (0.11,0.36)& 0.21 (0.15,0.27)&  0.25 (0.08,0.40)&  0.18 (0.09,0.28)& 0.21 (0.16,0.26)\\
 				&\textbf{Disease stage 3}&0.60 & 0.63 (0.46,0.83)& 0.68 (0.61,0.77)& 0.63 (0.58,0.68)& 0.63 (0.46,0.83)&  0.69 (0.61,0.77)&  0.63 (0.58,0.68)
 				\\
 			\hline	
 			\end{tabular}\hfill
 			%\end{centering}
               \vspace{0.5cm}
 		}
        {\raggedright \begin{small}\textit{* Observed stage-specific --5 stage-specific studies, Joint --5 stage-specific studies and 25 overall studies, and Ideal --30 stage-specific studies.} \end{small}\par}
 	\end{table}
    
\begin{figure}[h]
    \centering
    \includesvg[width=500pt]{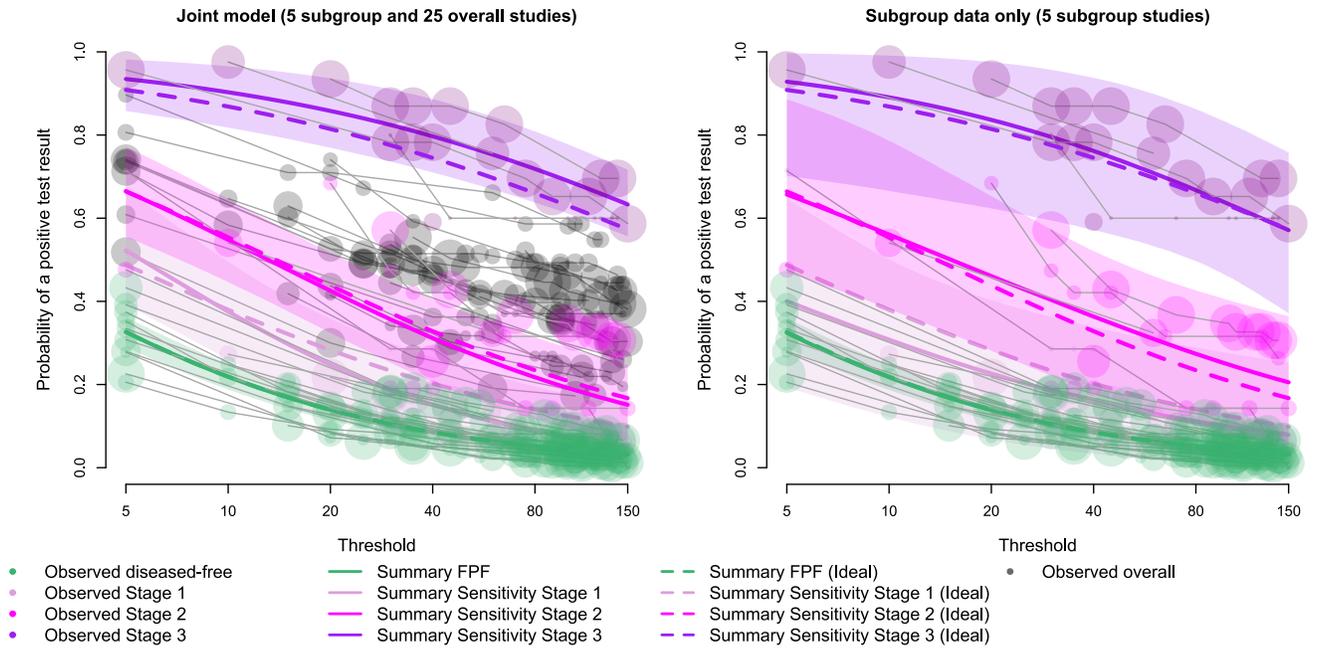}

    \caption{Results from models fitted to artificial data, Scenario II. Solid lines show results from fitting models to incomplete data. Dashed lines (identical across the two plots) show results from fitting models to the ideal full data, if it were available, and are shown for comparison. Shaded areas represent 95\% CrIs around the point estimates. The diameter of the observed points varies according to sample size. FPF=False positive fraction.}
    \label{fig:sims-cont-s2}
\end{figure}

For Scenario II, results are shown in Figure \ref{fig:sims-cont-s2} and Table \ref{tab:cont_sim}. Since in this case there was sufficient stage-specific data to estimate slopes for all stages, the differences between the two approaches are less striking compared with those from Scenario I. However, the gain in precision is still noticeable when all available information is used compared with using only the observed stage-specific data.

\subsection{Analysis of continuous test data from the motivating example}

We now apply this model to the AFP test from the motivating example. The available AFP data with some information regarding disease stage are summarized in Table \ref{tab:afp-data}. All 110 studies report data for the `no HCC' (disease-free) group. Of the three stages of disease, the largest amount of stage-specific data is for advanced stage HCC (15 studies). Only 6 studies reported sensitivity among individuals with early stage HCC. As described in Table \ref{tab:hcc-data}, an additional 51 studies reported estimates of AFP's sensitivity at one or more thresholds, alongside some information on HCC stage. Of these, 19 studies (labeled `Overall Type 1') in Table \ref{tab:afp-data}) reported overall data of the form described in Section \ref{sect:overall_cont}, i.e. overall sensitivity to detect HCC alongside proportion of HCCs that were at each stage. An additional 21 studies (labeled `Merged very early/early' in Table \ref{tab:afp-data}) provided sensitivity for the merged group of very early and early stage patients, as described in the case of binary tests in Section \ref{sect:binary_applic}, while 11 studies (`Overall Type 2' in Table \ref{tab:afp-data}) reported overall sensitivity alongside the proportion of HCCs there were advanced stage and the (merged) proportion that were very early or early stage. See the Supporting Material for more details. We note that, for this example, stage-specific data were reported at a maximum of 19 thresholds per study, while `overall' data were reported at a maximum of 5 thresholds per study (Table \ref{tab:afp-data}). \par
%A total of 110 studies and number of data points is 433.

	\begin{table}[!h]
 		\centering
 		\caption{\label{tab:afp-data}Summary of available AFP data}
 		%\begin{centering}
 		%\centering
 		\resizebox{\textwidth}{!}{%
 			\begin{tabular}{lrrr}
 				%\multicolumn{2}{c}{Continuous biomarkers} &
 				%\multicolumn{2}{c}{Imaging tests}&\multicolumn{2}{c}{Model based tests}&\multicolumn{2}{c}{Genomic biomarkers}\\  
 				\textbf{Data type}	&\textbf{No of studies }&\textbf{No of thresholds} &\textbf{Maximum no of thresholds per study }\\
 				\hline
                    \textbf{No HCC}  & 110 &  159& 48\\
 				\textbf{Very early stage}  & 7 &12  &6 \\
 				\textbf{Early stage} &  6& 43 &16 \\
 				\textbf{Advanced stage}& 15 & 32 & 19 \\
 				\textbf{Overall Type 1}& 19 & 15 & 4 \\
                    \textbf{Overall Type 2}& 11 & 15 & 5 \\
 				\textbf{Overlap very early/early}&  21&  19& 2
 				\\
 				\hline
 			\end{tabular}\hfill
 			%\end{centering}
 		}
 	\end{table}

\begin{figure}[h!]
    \centering
    \includesvg[width=510pt]{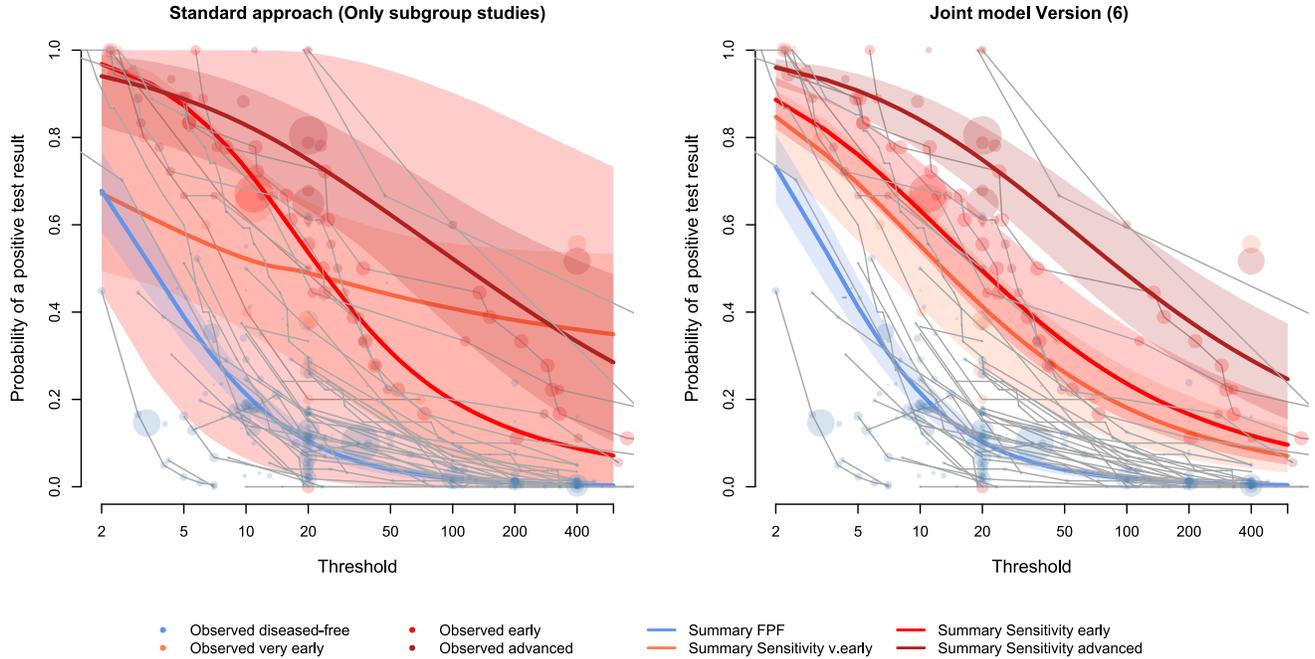}

    \caption{Pooled estimates of the probability of a positive test result for the AFP data from the HCC review, along with 95\% CrIs represented by the shaded areas. The diameter of the points varies according to sample size.}
    \label{fig:afp_results}
\end{figure}

We first fitted the \citet{jones2019quantifying} multiple thresholds model to data for each disease state separately, using data only from studies reporting stage-specific information. We assumed the Box-Cox family of transformations for $f()$ and assumed a $\mathrm{Uniform}(-3,3)$ prior for $\lambda.$ Figure \ref{fig:afp_results} (left) shows the pooled estimates, with 95\% CrIs, of the probability of a positive test result for each disease state, across the entire range of thresholds, based on this model. We notice that for most thresholds ($\geq 26$ ng/ml), the estimated sensitivity of AFP to detect very early stage HCC is higher than the estimated sensitivity to detect early stage HCC. This is a counterintuitive result based on clinical experts' opinion on this test, and is likely driven, similarly to the binary test results case, by the very small amount of stage-specific data --both number of studies, and number of thresholds within each study. The summary estimates are seen to be very imprecise, with largely overlapping CrIs between disease states, rendering results uninformative.  \par

Next, we fitted the joint model described in Section \ref{sect:overall_cont}, to both stage-specific and all forms of merged state data. We fitted several versions of the model, with different variance-covariance matrix structures (unrestricted or with particular restrictions) and pre-specifying $f() = log()$ or with $f()$ assumed to be in the Box-Cox transformation class with unknown parameter $\lambda$. We explored the comparative model fit, penalized for complexity, of these different versions of the model using the Deviance Information Criterion (DIC), which is defined as the sum of the residual deviance and the number of effective parameters pD \citep{spiegelhalter2002bayesian}. Details are available in the Supplementary Material. Of the models considered, the selected model based on the DIC was the version assuming a Box-Cox family transformation and correlations only among the location parameters of the four states (Version 4 in the Supplementary Material). Although the uncertainty around summary estimates was reduced, relative to results from the model fitted only to stage-specific data, the point estimates for sensitivity at an early stage were still lower than those for very early stage for thresholds $\geq 6.5$ ng/ml. For example, at a threshold of 20 ng/ml (the most commonly reported threshold), the pooled sensitivity at the very early stage was estimated to be 0.52 (95\% CrI 0.36,0.71) while  the pooled sensitivity at the early stage was estimated as 0.40 (95\% CrI 0.22,0.59). \par

As an additional exploratory step, we fitted the joint model for tests with binary results, introduced in Section \ref{sect:binary_full}, to the data relating to the threshold of 20 ng/ml only. The results are shown in Table \ref{tab:afp-res20}. The summary estimates for the very early and early stages were 0.36 (95\% CrI 0.19,0.54) and 0.43 (95\% CrI0.24,0.66) respectively. We see that the estimated sensitivity to detect `very early' stage HCC is quite different from that obtained from Version 4, and the illogical ordering of the summary sensitivities is not present in this analysis. \par 

Having observed these inconsistencies, and noting clinical opinion regarding the implausibility of the results from version 4, we re-fitted the continuous data model but additionally enforcing constraints, such that summary sensitivity must not decrease by stage. This was enforced through truncated priors for the $m_j$, $j=3,4$ location parameters, of disease stages 2 and 3, and by setting the parameters $m_6,m_7,m_8$, representing the slopes for the three disease stages, to be  equal, since their corresponding 95\% CrIs according to version 4 overlapped considerably. More specifically, while parameters $m_1,m_2,m_5,m_6$ (corresponding to the disease-free and stage 1 states), were assigned vague normal priors as described earlier, parameters $m_j$, $j=3,4$ were assigned truncated Normal distributions, taking values between $m_{j-1}$ and $\infty$. Note that similar constraints were also applied in \citet{dias2025bayesian} to correct for non-biologically possible observations arising due to small sample sizes. After enforcing these constraints (Version 6 in Supplementary Material), the residual deviance reduced by 1.1 points and pD increased by 0.4 points, resulting in a DIC reduced by 0.7 points compared with Version 4. The summary estimates are shown in Figure \ref{fig:afp_results} (right). The results at threshold 20 ng/ml  were reassuringly close to those based on the Binary model and are also summarized in Table \ref{tab:afp-res20}. \par

	\begin{table}[!h]
 		\centering
 		\caption{\label{tab:afp-res20}AFP test for detecting HCC: Point estimates and 95\% CrI comparison at threshold 20ng/ml of the probability of a positive test result.}
 		%\begin{centering}
 		%\centering
 		%\resizebox{\textwidth}{!}{%
 			\begin{tabular}{lrrr}
 				\multicolumn{1}{c}{} &
 				\multicolumn{1}{c}{Used only data at threshold 20ng/ml}&\multicolumn{2}{c}{Used data at all thresholds}\\ 
 				\textbf{State}	&\textbf{Joint model for binary tests}&\textbf{Version (4)} &\textbf{Version (6) }\\
 				\hline
                    \textbf{No HCC}  &0.11 (0.09,0.46)& 0.10 (0.08,0.13) &0.10 (0.08,0.12)   \\
 				\textbf{Very early stage}  & 0.36 (0.19,0.54)& 0.52 (0.36,0.71)& 0.41 (0.28,0.51) \\
 				\textbf{Early stage} & 0.43 (0.24,0.66)&0.40 (0.22,0.59) & 0.50 (0.39,0.62) \\
 				\textbf{Advanced stage}& 0.69 (0.57,0.78)&0.77 (0.69,0.86) & 0.75 (0.66,0.83) \\
 				
 				\hline
 			\end{tabular}\hfill
 			%\end{centering}
 		%}

 	\end{table}

\section{Discussion}\label{Sect:Concl}
 
Where disease stage-specific data are not available from all studies in a meta-analysis, making use of the more limited information available in other studies (`overall' or other merged-stage estimates of sensitivity, along with stage proportions) can improve the precision of pooled stage-specific accuracy estimates. We have demonstrated this with applications to both simulated and real data. This joint modeling approach may be particularly useful when dealing with continuous tests, where estimating the relationship between threshold and probability of testing positive in each disease state involves estimating more parameters.  As we have shown, including additional studies with stage proportions can help inform the required slope parameters.  \par

 The gains of this joint modeling approach may, in some of the cases we have explored, appear relatively underwhelming. This is because we are estimating multiple study-level parameters from aggregated data counts. For example, in including a study reporting only `overall' sensitivity in the joint model for tests with binary results, we are trying to estimate 3 parameters -- sensitivity for each disease stage -- from only 1 data point. This is only possible because of the random effects distributions over the stage-specific sensitivities. Precision gains are likely to vary considerably depending on the amount of between-study heterogeneity: we would expect greater precision gains when there is little to no heterogeneity across studies. In this case study, heterogeneity was estimated to be moderate/large for all the analyses.\par 

To explore the potential of the proposed approach, and as a proof of concept, we simulated a small number of data sets and fitted the models. This showed that incorporating merged stage data in the analysis can increase the precision of summary estimates and correct in some cases for biologically implausible results than may arise due to limited stage-specific data. A formal simulation study was beyond the scope of this paper, but could provide insights as to situations where incorporating merged stage data is most beneficial, e.g. by varying the amount of between-study heterogeneity and the number of studies reporting data of different types.\par 

A limitation of the models we have presented is that they do not utilize overall sensitivity reported in studies that do not additionally report any stage proportion data. This is because it would not be possible to estimate the additional study-specific parameters $p_{ij}$ from the single data point. We note that these studies could be incorporated, with potential for some learning, if an additional hierarchical model structure -- i.e. random effects distributions across studies -- were assumed for the stage proportions.  We considered this for our motivating data sets but observed that the reported proportions varied substantially across studies -- therefore keeping them as independent quantities felt more appropriate in this instance. \par

We note that our model produces estimates of test sensitivity conditional on a particular disease stage, but that other quantities may sometimes be of interest and can be calculated as functions of our summary parameters. For example, the sensitivity of a test to detect HCC at `very early or early' stage may be of clinical interest. Denoting summary sensitivity to detect stage $j$ by $S_j$, this quantity can be calculated as
$Pr(\text{test positive}|\text{stage 1 or stage 2} )=qS_{2}+(1-q)S_{3}$, where $q$ is the proportion of very early or early stage HCCs that are very early. To calculate  this, we would  need to assume a value for $q$, a function of the stage proportions. This could be based on routine data or a study population considered representative, or -- if extending the model to include a hierarchical structure for stage proportions $p_{ij}$, as described above -- could be calculated as a function of parameters of this distribution. \par 
 
As noted in the Introduction, our focus was different from  some other papers that have presented meta-analysis models for diagnostic test accuracy with multiple disease states. \citet{bipat2007multivariate} describe a model that focuses on the ability of the index test to correctly classify patients into different disease groups. This was applied to a case study assessing the accuracy of Endoluminal Ultrasonography for classifying patients into four stages of rectal cancer. Application of this model requires a full $J \times J$ classification table from each study, which are modeled with Multinomial likelihoods. \citet{bipat2007multivariate} assumed multivariate Normal random effects for the transformed probabilities of being correctly assigned to each disease group, similar to our approach. Additionally, \citet{bipat2010multivariate} extended this Multinomial model to account for the ordinal nature of the stage accuracy data.\par 

In contrast, our aim was to produce pooled estimates of the accuracy of tests in categorizing individuals into two groups (`diseased' and `disease-free'), conditional on true disease state $j = 0,...,J$. The data modeled are therefore $2\times (J+1)$ tables from each study rather than $J \times J$ tables. The most appropriate approach is likely to depend on the type of test and its possible use case. For example, AFP and US are used as first-line tests in screening for HCC in people with cirrhosis, and would always be followed-up with further investigations if positive. In this setting, the ability of these tests to classify individuals with HCC into the correct stage is of limited clinical relevance; rather, their performance in detecting the presence or absence of HCC is what is clinically meaningful.

Where clinical interest is instead in the ability of a test to classify patients into the correct disease stage, estimated using 
full $(J+1)\times (J+1)$ tables of results, inclusion of stage proportion data -- as explored in this paper -- could be similarly beneficial. \citet{corbett2020point} meta-analyzed the accuracy of point-of-care creatinine tests to classify patients into four categories of risk of developing post-contrast acute kidney injury. They used a similar approach to our model to allow the inclusion of studies reporting accuracy on overlapping categories, making use of available stage proportions: however, a fixed effects approach was adopted due to the small number of studies available. For the case of continuous tests our model could be extended too in a similar but slightly less straightforward way. An alternative approach to estimating stage-specific sensitivities when data are sparse could be to share parameters across different cancer types using a Bayesian shared parameter model \cite{dias2025bayesian} if such information is available.\par 
 
For the case of binary tests, we focused on extensions of the bivariate random effects meta-analysis model. However, there are situations where production of a summary Receiver Operating Characteristic (ROC) curve might be preferred. Several methods have been described for performing ROC analysis for multiple disease states for a single study \citep{zhou2014statistical,nakas2023roc}, but have not been extended into a meta-analysis context. The ideas described in this paper could be integrated with these approaches to extend the HSROC meta-analysis model \citep{rutter2001hierarchical} to multiple disease states, incorporating stage proportions. It is worth noting that summary ROC curves could be derived based on the model we have presented, drawing on the equivalence of the Bivariate and HSROC meta-analysis models \citep{harbord2007unification}.\par 

Finally, while our focus was on estimating test accuracy across multiple disease states, we note conceptual parallels to settings where accuracy is reported across other patient characteristics, 
 such as age group, gender or aetiology. For example, \citet{selby2019effect} aimed to produce meta-analytic summaries of test accuracy for different age groups. While some studies reported accuracy for the required groups, others reported only overall or merged accuracy alongside percentages of individuals in each age group. The ideas described in this paper could be applied to jointly model both types of data.

\section*{Acknowledgments}
HEJ and ED were supported by an MRC-NIHR New Investigator Research Grant (MR/T044594/1). This work was additionally supported by the NIHR Health Technology Assessment programme (project reference NIHR134670). 

\section*{Data availability}
The Data and R/jags code to reproduce some of the analyses in this paper are available on Git-hub at:  \href{https://github.com/FeniaDerezea/MDS_MA_DTA}{github.com/FeniaDerezea}

\bibliography{bibliography}

\end{document}

\maketitle
\begin{center}

	\begin{small}
		$1$. \textit{Population Health Sciences, Bristol Medical School, University of Bristol, UK}\\
		$2$. \textit{Manchester Centre for Health Economics, University of Manchester, UK}\\
	%	$2$.
	\end{small}	
	\end{center}
\section{Description of studies with collapsed disease data at multiple thresholds in the HCC review}
  In addition to the ideal subgroup data for the continuous test AFP in the motivating example, there were also 3 types of studies reporting some relevant information for some collapsed diseased state. In particular:
  \begin{itemize}
      \item \textbf{`Overall Type 1'}\\
            These are the standard overall studies described in the main text extensively for both tests with binary and continuous results. These studies report accuracy on the fully collapsed diseased state (i.e. patients could belong to any of the three diseases stages) along with the proportion of patients that belongs in each of the three disease stages (baseline proportions). See Section 4.1 in the main text for more details.
      \item \textbf{`Overlapping'}\\
            Similarly to the binary case, these studies reported accuracy for the overlapping disease group between the first two diseases stages (i.e `very early' and `early' HCCs for the motivating example). More precisely, each overlapping study $i$ reported the counts $x_{i1t}$, $x_{i1t}^{-}$, $xov_{it}=x_{i2t}+x_{i3t}$, $xov_{it}^{-}=x_{i2t}^{-}+x_{i3t}^{-}$, $x_{i4t}$, $x_{i4t}^{-}$ with $t$ varying depending on the study and state. Note that these studies do not provide any baseline proportions. 
      \item \textbf{`Overall Type 2'}\\
            These studies report accuracy on the fully collapsed disease state, similarly to the `Overall Type 1' studies. However, they instead report the proportion of participants in the overlapping `very early/ early' disease state, $qov_{i}=p_{i2}+p_{i3}$, along with the proportion of participants in the `advanced' state, $p_{i4}$. This type of studies report the following counts: $x_{i1t}$, $x_{i1t}^{-}$, $xall2_{it}$ and $xall2_{it}^{-}$.
  \end{itemize}
  
  \section{Model specification of studies with stage information on collapsed disease groups at multiple thresholds in the HCC review}
  To jointly model all types of studies, for the AFP test from the motivating example, we follow the steps described in Section 4.3 in the main text.
  \begin{itemize}
      \item \textbf{`Overall Type 1'}\\
      These are the standard overall studies and are modeled as described in Section 4.3 in the main text.
      \item \textbf{`Overlapping'}\\
      We can model the ``Overlap very early/early" studies following the same approach as for the binary tests. In particular, if we denote with $xov_{it}$ the counts of the overlapping group of study $i$ and threshold $t$, we can assume that these are generated according to a series of conditional Binomial distributions as
$$xov_{i1}\sim Bin(Sov_{i1},N_{i2}+N_{i3})$$
$$xov_{it}\sim Bin(\frac{Sov_{it}}{Sov_{it-1}},xov_{it-1}),\quad t=2,\dots,Tov_{i}$$
where $Tov_{i}$ is the number of reported thresholds in the overlapping state, and $Sov_{it}$ is the overlapping sensitivity at each threshold $t$, which is defined as:
$$Sov_{it}=q_{i}S_{i2t}+(1-q_i)S_{i3t}$$
The unknown proportion of very early patients within the overlapping group, $q_i$, is defined as shown in Section 3.3 and modeled in the same way. 
      \item \textbf{`Overall Type 2'}\\
       We assume that the corresponding $xall2_{it}$ counts are generated as:
$$xall2_{i1}\sim Bin(Sall2_{i1},N_{i2}+N_{i3}+N_{i4})$$
$$xall2_{it}\sim Bin(\frac{Sall2_{it}}{Sall2_{it-1}},xall2_{it-1}),\quad t=2,\dots,Tall2_{i}$$
and the overall sensitivities $Sall2_{it}$ in this case are given by
$$Sall2_{it}=qov_{i}Sov_{it}+p_{i3}S_{i3t}$$
where $qov_{i}$ are known proportions of patients in the overlapping group.
  \end{itemize}
  \section{Additional results from the application to the AFP dataset}
  In Section 6.2 we apply the proposed joint model for tests with continuous results to the AFP test from the motivating example. We first fit the model in its most general form, with an  $8\times 8$ unrestricted covariance matrix for the random effects and unknown Box-Cox parameter, $\lambda$ (Version 1). Estimates with 95\% CrIs for the model parameters, along with goodness of fit statistics, are summarized in Table \ref{tab:afp-gof}. Using these results as a guide, we fit some further simplified versions of the model considering different structures for the variance-covariance matrix $\Sigma$. In particular, we first assume that there are no correlations between the scale and location parameters (Version 2), then assume additionally that there are no dependencies between the scale parameters (Version 4), and finally we assume full independence among all parameters (Version 5). Since $\lambda$ was estimated to be relatively small in magnitude, we also fit a version where $\lambda =0$, effectively using a log-link (Version 3). \par 
	\begin{table}[!h]
 		\centering
 		\caption{\label{tab:afp-gof}Model fit comparison for the AFP data}
 		%\begin{centering}
 		%\centering
 		\resizebox{\textwidth}{!}{%
 			\begin{tabular}{lrrrrrr}
 				%\multicolumn{2}{c}{Continuous biomarkers} &
 				%\multicolumn{2}{c}{Imaging tests}&\multicolumn{2}{c}{Model based tests}&\multicolumn{2}{c}{Genomic biomarkers}\\  
 				\textbf{ }	&\textbf{(1) 8D Box-Cox } &\textbf{(2) 4Dm,4Ds Box-Cox} &\textbf{(3) 4Dm,4Ds Log}& \textbf{(4) 4Dm,Ind s Box-Cox}&\textbf{(5) Independence Box-cox}& \textbf{(6) 4Dm,Ind s Box-Cox (force)} \\
 				\hline
                    \textbf{Resdev}  &521.7  &  521.1& 526.1&524.2 &524.0&523.1\\
 				\textbf{pD}  & 188.7 &  190.6& 190.7&188.1  &192.6&188.5\\
 				\textbf{DIC} & 710.4&    711.7& 716.8& 712.3 &716.6&711.6\\
                    \hline
                    \textbf{$m_{1}$}& 1.22 (1.00,1.39)  & 1.25 (1.06,1.42) & 1.23 (0.99,1.46) &1.25 (1.07,1.42) &1.23 (1.03,1.39)&  1.26 (1.08,1.42)\\
                    \textbf{$m_2$}& 2.87 (1.93,3.80)   &2.66 (1.77,3.80) & 3.23 (1.99,4.75)&2.65 (1.70,3.76) &2.50 (1.07,4.28)&  2.21 (1.69,2.66)\\
                    \textbf{$m_3$}& 2.13 (1.60,2.83)   &2.29 (1.67,3.07) &2.64 (1.84,3.66)&2.23 (1.61,2.96) &2.21 (1.59,3.02)&  2.51 (2.11,3.11)\\
                    \textbf{$m_4$}&  3.79 (3.16,4.64)&  3.85 (3.20,4.76) &4.93 (4.21,5.83)&3.76 (3.13,4.63) &4.02 (3.28,5.10)&  3.52 (3.01,4.22)\\
                    \textbf{$m_5$}& -0.51 (-0.73,-0.26)  & -0.49 (-0.71,-0.26)& -0.18 (-0.32,-0.04)& -0.51 (-0.74,-0.28) &-0.45 (-0.69,-0.20)&  -0.53 (-0.74,-0.31)\\
                    \textbf{$m_6$}& 0.14 (-0.50,0.99)   & 0.26 (-0.43,1.17)& 0.58 (-0.13,1.57)& 0.27 (-0.43,2.07) &0.70 (-0.26,2.83)&  -0.11 (-0.38,0.26)\\
                    \textbf{$m_7$}& -0.34 (-0.82,0.21)  & -0.26 (-0.77,0.48)& 0.04 (-0.46,0.85)& -0.30 (-0.68,0.45) &-0.25 (-0.65,0.42)&  -0.11 (-0.38,0.26)\\
                    \textbf{$m_8$}&  0.01 (-0.44,0.46)  & 0.06 (-0.38,0.51)& 0.47 (0.11,0.84)& -0.02 (-0.42,0.43) &0.08 (-0.35,0.58)&  -0.11 (-0.38,0.26)\\
                    \textbf{$\lambda$}& -0.11 (-0.18,-0.05)  &-0.10  (-0.16,-0.04)& 0& -0.11 (-0.17,-0.04)&-0.10 (-0.16,-0.03)&  -0.12 (-0.17,-0.06)\\
 				\\
 				
 			\end{tabular}\hfill
 			%\end{centering}
 		}
 	\end{table}
    We compare the different versions based on their goodness of fit using the Deviance Information Criteria, which is defined as the sum of the residual deviance and the number of effective parameters pD. Note that JAGS by default returns a different value for the number of effective parameters (pV), which can sometimes be unstable. For this reason, we calculated pD ``by hand" outside of JAGS in R. Lower values of DIC are an indication of a better fit. We see that the DIC is similar for versions 1,2 and 4. Versions 3 and 5 assuming a log link and full independence respectively, returned higher DIC values. Based on these results and following the rule of thumb according to which we select a more complicated model if DIC reduces by 3 or more, Version 4 seems like a reasonable choice.\par 
    Finally because of the implausibility of the results produced according to version 4 we re-fit this model constraining sensitivity to not get smaller as disease stage increases (Version 6).